\documentclass{edm_template}

\usepackage{color}
\usepackage[normalem]{ulem}
\usepackage[latin1]{inputenc}
\usepackage{amssymb,amsmath,array}
\usepackage{comment}
\usepackage{url}
\usepackage{cite}
\usepackage{enumitem}
\usepackage{textcomp}
\usepackage{booktabs}
\usepackage{tikz}
\usepackage{pgfplots}
\usepackage{rotating}

\usetikzlibrary{3d,decorations.text,shapes.arrows,positioning,fit,backgrounds}
\usepgfplotslibrary{groupplots}
\usepgfplotslibrary{fillbetween}
\definecolor{colDefault}{RGB}{50,136,189}
\definecolor{colDefaultFill}{RGB}{50,136,189}

\colorlet{colCluster}{colDefault}
\colorlet{colClusterError}{colDefault}

\colorlet{colEncode}{colDefault}
\definecolor{colCompare}{RGB}{252,141,89}

\colorlet{colCluster0}{colCluster}
\definecolor{colCluster1}{RGB}{213,62,79}
\definecolor{colCluster2}{RGB}{252,141,89}
\definecolor{colCluster3}{RGB}{254,224,139}
\definecolor{colCluster4}{RGB}{153,213,148}
\definecolor{colClusterRandom}{RGB}{0,0,0}

\pgfplotsset{
    every axis/.style={
      scaled x ticks=false,
      scaled y ticks=false,
    },
}

\pgfplotsset{
    colormap={simcolormap}{
        rgb255=(50,136,189),
        rgb255=(153,213,148),
        rgb255=(254,224,139),
        rgb255=(252,141,89),
        rgb255=(213,62,79),
    },
}

\pgfplotsset{
    FullClusterPlot/.style={
      width=\linewidth,
      height=0.95\linewidth,
      xmin=0,xmax=30,
      ymin=0,ymax=1,
      ylabel style={
             font=\footnotesize,
      },
      xlabel style={
             font=\footnotesize,
      },
      xlabel={Months},
      ylabel={Similarity},
      every axis plot/.append style={line width=2.0pt},
      legend style={at={(axis cs:15,0.03)},anchor=south, legend columns=3},
      xtick={0,5,10,15,20,25,30},
      grid=both,
      grid style={line width=.2pt,draw=gray!50},
    },
    ClusterPlot/.style={
      scale only axis,
      width=0.2\textwidth,
      height=4.5cm,
      xmin=0,xmax=30,
      ymin=0,ymax=1,
      ylabel style={
             font=\footnotesize,
      },
      Cluster/.style={color=colCluster,mark=none, line width=2.0pt},
      ClusterError/.style={color=colClusterError,mark=none,opacity=0.3},
      legend style={at={(axis cs:15,0.05)},anchor=south, legend columns=-1},
      xtick={0,5,10,15,20,25,30},
      grid=both,
      grid style={line width=.2pt,draw=gray!50},
      enlargelimits = false,
    },
    ClusterPlotVocab/.style={
      scale only axis,
      width=0.2\textwidth,
      height=1.7cm,
      ylabel style={
             font=\footnotesize,
      },
      xmin=0,xmax=30,
      ymin=10.3,ymax=11.5,
      ylabel={},
      Data/.style={color=colCluster,mark=none, line width=1.0pt},
      xticklabels={,,},
      xtick={0,5,10,15,20,25,30},
      grid=both,
      grid style={line width=.2pt,draw=gray!50},
      enlargelimits = false,
    },
    ClusterPlotPhrases/.style={
      scale only axis,
      width=0.2\textwidth,
      height=1.7cm,
      ylabel style={
             font=\footnotesize,
      },
      xticklabels={,,},
      xmin=0,xmax=30,
      ymin=2.5,ymax=3.6,
      ylabel={},
      Noun/.style={color=colCluster,mark=none, line width=1.0pt},
      Verb/.style={color=colCluster,mark=none, line width=1.0pt, dashed},
      xtick={0,5,10,15,20,25,30},
      grid=both,
      grid style={line width=.2pt,draw=gray!50},
      enlargelimits = false,
    },
    ClusterPlotWords/.style={
      scale only axis,
      width=0.2\textwidth,
      height=1.7cm,
      ylabel style={
             font=\footnotesize,
      },
      xmin=0,xmax=30,
      ymin=700,ymax=1300,
      ylabel={},
      Data/.style={color=colCluster,mark=none, line width=1.0pt},
      xtick={0,5,10,15,20,25,30},
      grid=both,
      grid style={line width=.2pt,draw=gray!50},
      enlargelimits = false,
    },
}

\newcommand{\set}[1]{\left\{#1\right\}}

\newcommand{\txt}{\ensuremath{t}}
\newcommand{\student}{\ensuremath{\alpha}}
\newcommand{\students}{\ensuremath{\mathcal{A}}}
\newcommand{\otherstudent}{\ensuremath{\beta}}
\newcommand{\txts}{\ensuremath{T}}
\newcommand{\txtsstud}{\ensuremath{\txts_\student}}
\newcommand{\txtsstudi}[1]{\ensuremath{\txts_{#1}}}

\newcommand{\simfuncname}{\ensuremath{s}}
\newcommand{\simfunc}[2]{\ensuremath{\simfuncname(#1,#2)}}

\newcommand{\analset}{\ensuremath{T_{analyze}}}
\newcommand{\networkset}{\ensuremath{T_{network}}}
\newcommand{\trainset}{\ensuremath{T_{train}}}

\newcommand{\valset}{\ensuremath{T_{val}}}

\newcommand{\profile}[1]{\ensuremath{P_{#1}}}
\newcommand{\aprofile}[1]{\ensuremath{\hat P_{#1}}}
\newcommand{\centroid}{\ensuremath{C}}

\newcommand{\psim}{\ensuremath{p}}
\newcommand{\ptime}{\ensuremath{\tau}}

\newcommand{\Sim}{\textsc{Sim}}

\newcommand{\cluster}[1]{\ensuremath{\mathcal{C}_{#1}}}

\newcommand{\clustererror}{\ensuremath{E_{\mathcal{C}}}}

\newcommand{\secref}[1]{Section~\ref{sec:#1}}
\newcommand{\figref}[1]{Figure~\ref{fig:#1}}
\newcommand{\tabref}[1]{Table~\ref{tab:#1}}

\begin{document}

\title{Investigating Writing Style Development in High School}

\numberofauthors{3} 
\author{
\alignauthor
Stephan Lorenzen\\
       \affaddr{University of Copenhagen}\\
       \email{lorenzen@di.ku.dk}
\alignauthor
Niklas Hjuler \\
       \affaddr{University of Copenhagen}\\
       \email{Hjuler@di.ku.dk}
\alignauthor
Stephen Alstrup \\
       \affaddr{University of Copenhagen}\\
       \email{alstrup@di.ku.dk}
}


\maketitle
\begin{abstract}
In this paper we do the first large scale analysis of writing style development among Danish high school students. More than 10K students with more than 100K essays are analyzed. Writing style itself is often studied in the natural language processing community, but usually with the goal of verifying authorship, assessing quality or popularity, or other kinds of predictions.

In this work, we analyze writing style changes over time, with the goal of detecting global development trends among students, and identifying at-risk students. We train a Siamese neural network to compute the similarity between two texts. Using this similarity measure, a student's newer essays are compared to their first essays, and a writing style development profile is constructed for the student. We cluster these student profiles and analyze the resulting clusters in order to detect general development patterns.
We evaluate clusters with respect to writing style quality indicators, and identify optimal clusters, showing significant improvement in writing style, while also observing suboptimal clusters, exhibiting periods of limited development and even setbacks. 

Furthermore, we identify general development trends between high school students, showing that as students progress through high school, their writing style deviates, leaving students less similar when they finish high school, than when they start.
\end{abstract}


\section{Introduction}
One of the most essential skills, learned during the course of primary, secondary and high school,
is writing. While the main focus of primary school are on basic writing skills (such as grammar), secondary or high school will be more focused on improving \emph{the linguistic writing style} of a student, that is, the quality of the written text as perceived by the reader. With many jobs being highly dependent on producing relatively large amounts of well-written text, no justification is needed for why \emph{good writing} is an essential skill.

The definition of quality in linguistic writing style is widely discussed \cite{spandel2001creating,growthinenglish}. While correct grammar being a prerequisite, several other measures are also correlated to writing style being perceived as good, for instance use of vocabulary, sentence structure and readability \cite{Pitler2008RevisitingRA}.
Our focus in this work will mainly be writing style \emph{development} through the course of high school, while writing style quality will have a secondary role. We consider data from Danish high schools, consisting of Danish essays, and investigate the general development patterns among the students during the three years of study.
The end goal is to be able to provide feedback to teachers about the development of their students' writing styles.
We identify patterns among thousands of students across different classes and institutions, allowing us to provide teachers with new insights, which the data available to the teacher might not show. For instance insights about students, whose writing style development patterns may be unique within their own classes.

By itself, our method potentially allows for identifying students with deviating writing styles development (which might be good or bad), or students with sudden significant changes in writing style, which could be an indicator of cheating.
However, we also consider several measures for the \emph{quality} of writing. We investigate how these measures correlate with the different patterns of writing style development found, as a mean to detect optimal and suboptimal development profiles with respect to text quality. Information of this kind could be used to help teachers tailor their teaching style to specific groups of students, who may need training in specific areas challenging to their development profile.

\subsection{Our Contribution}
As mentioned, we concern ourselves with the development of linguistic writing style (as opposed to e.g. handwriting) during the course of high school. Specifically, we investigate the development of writing style in Danish essays handed-in by students in Danish high schools
\footnote{Note, that high school in Denmark usually consists of three years of study with students normally starting at age 15-17 and finishing at age 18-20.}.

We are interested in determining general patterns of development, and to discuss which of the patterns are optimal, in the sense of improving writing style quality. In particular, we consider the following questions: 
\begin{itemize}
    \item How does the writing style of a student develop, and what are the typical kinds of development in writing style?
    \item How does writing style changes correlate with measures of quality?
    \item How does writing style similarity between students behave, with respect to how far the students are in their education?
\end{itemize}

Our study is based on data from the company MaCom\footnote{The data set is proprietary and not publicly available}, who is behind the learning management system Lectio, a system used by 90\% of Danish high schools. Students submit their written essays through Lectio, giving MaCom access to a huge corpus of Danish texts by high school students, marked with author and date of submission.

Our approach is based on methods from authorship verification; in order to learn a similarity measure for writing style, we consider examples of writing styles in texts from the same or different authors, similar to how it is done in verification tasks.
We use a Siamese neural network for learning this similarity measure. While training, time is not taken into account. Assuming that writing style actually changes over time, this will lead to a suboptimal network. However, testing the network, we see clear patterns in how the "errors" distribute for a single author, indicating that the network simulates the best similarity measure possible, and the "errors" are actual changes in writing style.
Using this method, writing style development profiles are generated and clustered for a large set of students. Analyzing the clusters, we see optimal and suboptimal types of development. In general, the average similarity is found to decay with time to a great extend, which corresponds well with the general perception, that writing style changes during high school, and also matches conclusions made in the literature \cite{ESANN19Ghostwriter,hansen2014,changeovertime}.

While this paper presents a case study of the data from MaCom, the methods used for analysis are of independent interest, and not specific to the Danish language or high school, except for the neural network, which would at least require retraining in the given language. Considering other network architectures than the one used in this work, might also improve upon the analysis, see for instance \cite{Qian2018} for a network used with English.

\subsection{Related Work}
Writing style analysis, in one way or another, has been studied in the natural language community for many years. Typically, the analysis of writing style is used as a middle link for tasks such as \emph{authorship verification} \cite{ESANN19Ghostwriter,Qian2018,stamatos2009}, in which a text of unknown authorship is given, together with a set of texts by some known author, and we wish to verify, whether the given author is the author of the unknown text. Similarly, in \emph{authorship attribution} the unknown text must be attributed to one of several known authors. Traditional methods for verification and attribution utilize both unsupervised methods from the field of outlier detection \cite{stamatos2009}, as well as standard supervised learning techniques, such as SVMs \cite{hansen2014} and techniques based on neural networks \cite{ESANN19Ghostwriter, Qian2018}.

Other uses of writing style analysis include distinguishing features of the writer (e.g. sex and age \cite{ageandgender, gender,ageandgender2,gender2}, demographics \cite{DemoPersonalityCUltural}, or nationality \cite{nationality1}), using supervised learning algorithms such as SVMs, random forest, and neural networks.
Other studies have investigated written conversations on online forums, trying to infer whether one person is trying to convince another \cite{isheargueing}.

Some studies investigate the quality of writing, for instance prediction of popularity of news articles \cite{Newspopularity1}, or the quality of scientific articles \cite{TACL76}. 
The former uses the popularity of an article on social media as a measure of quality,
while the quality measure of scientific papers considered in the latter
is based on acceptance of a paper to "The Best American Science Writing", an anthology of popular science articles published in the United States on a yearly basis.

Few studies consider development of writing style as the main objective.
\cite{handwriting} uses neural network models to track style of \emph{handwriting} (i.e. not linguistic writing style) and
investigate the development of handwriting among young students, and how similar it is when compared to different students, in the same/different grade level.
\cite{growthinenglish} shows how students in higher grades get higher scores for their essays from teachers, in a blind experiment, where all student information is hidden from the grading teacher. \cite{changeovertime} considers two famous Turkish writers, investigating their change in writing style over time, the most significant finding being average word length increasing with the age of the author.

Finally, several studies related to writing style have been conducted using the data available from MaCom. \cite{hansen2014} investigates temporal aspects of authorship attribution, and concludes that considering more recent essays improves authorship attribution algorithms, indicating that the writing style among high school students does indeed change with time.
\cite{ESANN19Ghostwriter} also uses the MaCom data for testing their neural network based authorship verification methods; their results also support these findings.



\section{Methods and Setup}\label{sec:method}
This section describes our experimental setup and methods. We start by giving some basic notation.

We consider a set of students \students, and let $\student\in\students$ denote a single student with texts $\txt\in\txtsstud$. 
Furthermore, let $\txts = \cup_{\student\in\students} \txtsstud$ denote the entire corpus of texts.

Since our main focus is how the writing style of a student develops during the time they spend in high school, we are interested in computing a similarity function $\simfuncname: \txts \times \txts \to [0,1]$, allowing us to compare the writing style between two texts. As mentioned, we utilize a Siamese neural network to compute \simfuncname; this approach is widely used for computing writing style similarity \cite{Qian2018,ESANN19Ghostwriter,siamese1}, specifically, our network will be similar to that of \cite{ESANN19Ghostwriter}. \secref{network} will describe our network in detail.

The similarity measure \simfuncname\ found will then be utilized for writing style analysis. Primarily, we will focus on determining development patterns by generating a \emph{writing style development profile} $\profile{\student}$ for each student \student. These profiles are then clustered and analyzed with respect to different measures for text quality. The profile generation and clustering are described in more detail in \secref{clustering}.

Finally, we also explore how the similarity between random students change depending on their current progress through high school. This is done by sampling random pairs of texts $\txt_1\in\txtsstudi{\student}, \txt_2\in\txtsstudi{\otherstudent}$ and computing their similarity. We then consider how the similarity changes depending on if \student\ and \otherstudent\ are in the same grade or not.

\begin{figure}[t]
    \centering
    \resizebox{0.42\textwidth}{!}{
        \begin{tikzpicture}[
  node distance = 10mm,
  every node/.style = {
    font=\tiny
  },
  einput/.style = {
    rectangle,
    draw=black,
    font=\footnotesize,
    thick,
    fill=white,
    minimum width=1.7em,
    minimum height=1.7em
  },
  eembd/.style = {
    rectangle,
    draw=gray!20,
    font=\footnotesize,
    thick,
    fill=gray!20,
    minimum width=7.2em,
    minimum height=2em
  },
  econv/.style = {
    rectangle,
    draw=gray!20,
    font=\scriptsize,
    thick,
    fill=gray!20,
    minimum width=3.4em,
    minimum height=2em
  },
  egmp/.style = {
    rectangle,
    draw=gray!20,
    thick,
    fill=gray!20,
    minimum width=3.4em,
    minimum height=2.1em
  },
  emerge/.style = {
    rectangle,
    draw=gray!20,
    thick,
    fill=gray!20,
    minimum width=8em,
    minimum height=2em
  },
  edense/.style = {
    rectangle,
    draw=gray!10,
    font=\footnotesize,
    thick,
    fill=gray!20,
    minimum width=6em,
    minimum height=2.5em
  },
  eout/.style = {
    circle,
    draw=black,
    thick,
    fill=black,
    inner sep=0pt,
    minimum width=0.7em,
    minimum height=0.7em
  },
  cedge/.style = {
    ->,
    draw=gray!70,
    thick
  }
]

\node[einput] (input1) at (-1.4,0) {$\txt_1$};
\node[einput] (input2) at (1.4,0) {$\txt_2$};

  \foreach \X in {1,2} {
    \node[eembd, text width=5em]  (embd\X) [below = 1.5em of input\X .south] {\tiny{\textbf{EMBEDDING}} \tiny{$d$=$5$}};
    \node[econv, text width=2.7em]  (conv\X a) [below left = 0.7em and 0.2em of embd\X .south]
         {\tiny \textbf{CONV.} \tiny $k$=$8$,\ \ \ $n$=$700$};
    \node[econv, text width=2.7em]  (conv\X b) [below right = 0.7em and 0.2em of embd\X .south]
         {\tiny \textbf{CONV.} \tiny $k$=$4$,\ \ \ $n$=$500$};
    \node[egmp]  (gmp\X a) [below = 0.7em of conv\X a.south] {\textbf{GMP}};
    \node[egmp]  (gmp\X b) [below = 0.7em of conv\X b.south] {\textbf{GMP}};
    \draw[cedge] (input\X) -> (embd\X);
    \draw[cedge] ([xshift=-1em] embd\X .south) -> (conv\X a.north);
    \draw[cedge] ([xshift=1em] embd\X .south) -> (conv\X b.north);
    \draw[cedge] (conv\X a) -> (gmp\X a);
    \draw[cedge] (conv\X b) -> (gmp\X b);
  }
  
\node[emerge] (merge) at (0,-14.5em) {\textbf{MERGE}};

\draw[cedge] (gmp1a.south) -> ([xshift=-3em] merge.north);
\draw[cedge] (gmp1b.south) -> ([xshift=-1em] merge.north);
\draw[cedge] (gmp2a.south) -> ([xshift=1em] merge.north);
\draw[cedge] (gmp2b.south) -> ([xshift=3em] merge.north);

\node[edense, text width=4em, align=center] (dense) [below = 1em of merge.south] {\tiny \textbf{DENSE} $4\times 500$};
  
\draw[cedge] (merge.south) -> (dense.north);

\node[eout] (out1) [below left = 1em and 1.8em of dense.south] {};
\node[eout] (out2) [below right = 1em and 1.8em of dense.south] {};
\node[] (outl1) [below = 0.0em of out1.south] {\tiny $\simfunc{\txt_1}{\txt_2}$};
\node[] (outl2) [below = 0.0em of out2.south] {\tiny $1$-$\simfunc{\txt_1}{\txt_2}$};

\draw[cedge] ([xshift=-2.1em] dense.south) -> (out1.north);
\draw[cedge] ([xshift=2.1em] dense.south) -> (out2.north);

\begin{scope}[on background layer]
  \node[draw=colEncode,thick,fit=(embd1) (gmp1b)] (encode1) {};
  \node[draw=colEncode,thick,fit=(embd2) (gmp2b)] (encode2) {};
  \draw[thick,draw=colEncode] (encode1) -- (encode2);
  \node[colEncode] [right = -0.1em of encode2.east, rotate=-90, anchor=south] {\textsc{Encoding}};
  \node[draw=colCompare,thick,fit=(merge) (outl2)] (comparison) {};
  \node[colCompare] [right = -0.1em of comparison.east, rotate=-90, anchor=south] {\textsc{Comparison}};
\end{scope}

\end{tikzpicture}
    }
    \caption{Network architecture.}\label{fig:network}
\end{figure}
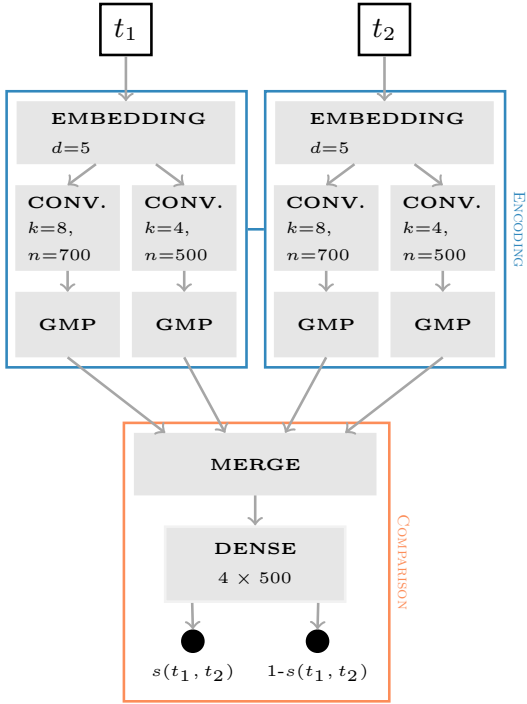

\subsection{Text Similarity using a Siamese Neural Network}\label{sec:network}
As mentioned, we use a Siamese neural network for computing the similarity $\simfunc{\txt_1}{\txt_2}$ between two texts $\txt_1$ and $\txt_2$. We considered several different architectures, using different input channels (e.g. char, word, part of speech tags). These architectures were evaluated using a validation set (see \secref{data}), and the best architecture was selected, as shown in \figref{network}. The network relies only on character level inputs.

The basic philosophy behind the network is to a) \emph{encode} the two texts in some space using a replicated encoder network with shared weights, and b) \emph{compare} the two texts in this space.
\begin{itemize}
    \item The encoder network in the \textbf{encoding} part of our network consists of a character embedding (using ReLu activation functions), followed by two different convolutional layers (\textbf{CONV}): one using kernel size $k=8$ and $n=700$ filters, and one using $k=4$ and $n=500$, each followed by global max pooling layers (\textbf{GMP}).
    \item In the \textbf{comparison} part of the network, the \textbf{MERGE} layer first computes the absolute difference between the outputs of the two encoder networks. Afterwards, four dense layers (\textbf{DENSE}) with 500 neurons each are applied, using ReLu for activation function and with a dropout of 0.3. Finally a two neuron softmax layer is used to normalize the output.
\end{itemize}
Using the convolutional layers, the network extracts character n-grams. Specifically, it compares 8- and 4-grams. Character n-grams have been shown to be an important feature in writing style analysis tasks such as authorship attribution \cite{stamatos2009}. We did also consider architectures using recurrent networks, however none of them performed as well as convolutional networks.

\subsection{Student Profiling and Clustering}\label{sec:clustering}
As mentioned, we construct writing style development profiles for the students, in order to analyze the general development patterns. The profile \profile{\student}\ for student \student\ is constructed by first determining their initial writing style.
The natural way to do so, and indeed our approach, is to consider their early work. One or more texts may be used to represent the initial writing style, as a trade off between the amount of data available for the profile and the robustness of the initial writing style estimation.
$\profile{\student}$ then consists of a chronologically ordered sequence of similarities, between any $\txt\in\txtsstud$ and this initial writing style. More specifically, if $\txt_1,\txt_2,...,\txt_{|\txtsstud|}$, $\txt_i\in\txtsstud$ is a chronologically ordering of \txtsstud, we compute the similarity $\psim_i$ between $\txt_i$ and the initial writing style by:
\begin{align*}
    \psim_i = \frac{1}{m}\sum_{j=1}^m \simfunc{\txt_i}{\txt_j},
\end{align*}
where $m$ is the number of texts used for representing the initial writing style. Since the first $m$ texts are part of the initial writing style, $\psim_1,\psim_2,...,\psim_m$ are not independent, and thus we exclude the first $m-1$ texts, and re-index such that $\psim_j = \psim_{i-m+1}$. Furthermore, for each text, we let $\ptime_j$ denote the time in months since $\txt_m$ was written, i.e. the time since $\psim_0$, with $\ptime_0 = 0$. Now, the final profile becomes the sequence consisting of pairs $(\ptime_j, \psim_j)$ of length $|\txtsstud|-m+1$. Note that the profile now describes a curve.

These profiles are now clustered using a slightly modified $k$-means clustering.
Before clustering, for each profile $\profile{\student}$, an approximate profile $\aprofile{\student}$ is constructed by interpolating values between any two consecutive pairs $(\ptime_j,\psim_j)$ and $(\ptime_{j+1},\psim_{j+1})$,
in intervals of 0.05 months. Thus $\aprofile{\student}$ becomes a vector $\aprofile{\student} \in [0,1]^{\ell_\student}$ consisting of similarities for every 0.05 month, with length $\ell_\student$.


These approximate profiles are then clustered. The clustering is complicated by profiles having variable length: $\aprofile{\student}$ has length $\ell_\student$ depending on $\ptime_{|\txtsstud|-m+1}$ (the time span between $\txt_m$ and $\txt_{\txtsstud}$), specific to $\student$.
Hence, distance computation used in the clustering algorithm is modified slightly; we compute the distance $dist(\aprofile{\student},\aprofile{\otherstudent})$ between two profiles $\aprofile{\student}$ and $\aprofile{\otherstudent}$ by computing the Euclidean distance between the prefixes of length $\ell = \min\set{\ell_\student,\ell_\otherstudent}$ of the two profiles:
\begin{align*}
    dist(\aprofile{\student},\aprofile{\otherstudent}) = dist_E(\aprofile{\student}[1...\ell],\aprofile{\otherstudent}[1...\ell]),
\end{align*}
where $dist_E$ denotes the Euclidean distance, and $v[1...n]$ denotes the prefix of length $n$ of vector $v$.

Similarly, when computing centroid $\centroid_r$ for cluster $\cluster{r}$, profile $\aprofile{\student}$ contributes only to the $\ell_\student$ first entries of $\centroid_r$. Thus, with $\cluster{r}^j = \set{\aprofile{\student} | \aprofile{\student} \in \cluster{r}, \ell_\student \leq j}$, the $j$'th entry of $\centroid_{r}$ is then computed as:
\begin{align*}
    \centroid_{r}[j] = \frac{1}{|\cluster{r}^j|} \sum_{\aprofile{\student} \in \cluster{r}^j} \aprofile{\student}[j]
\end{align*}
where $v[j]$ denotes the $j$'th entry of vector $v$.

The clustering is initiated by selecting $k$ profiles at random as the initial clusters, and then continually reassigning profiles and recomputing centroids for clusters. Having reassigned the profiles, the $\clustererror$ is computed:
\begin{align*}
    \clustererror = \frac{1}{|\students|}\sum_{r=1}^{k} \sum_{\student\in\cluster{r}} dist(\aprofile{\student}, \centroid_r)
\end{align*}
The algorithm iterates until the change in cluster error $\clustererror$ is sufficiently small ($\clustererror \leq 10^{-6}$), or until a set number of maximum iterations (100) is reached.

Selecting the number of clusters $k$ is an inherent problem in all unsupervised learning task. One approach is to base the decision on domain knowledge, and select the "right" number of clusters. We will instead make use of
the so called \emph{elbow heuristic} which relies on looking at how the error decreases with the number of cluster and pick at the "elbow"
in the resulting curve \cite{elbow}.

Having determined the parameter $k$ and found $k$ clusters, we compute a few statistics and writing quality indicators for each cluster. Specifically, we will compute the average \emph{noun and verb phrases}, defined as the ratio between nouns and sentences, and the ratio between main verbs and sentences respectively. These measures, especially verb phrases, have been shown to correlate well with readability, which correlates with text quality \cite{Pitler2008RevisitingRA}. Furthermore, we compute the \emph{simple measure of Gobbledygook} (SMOG) grade \cite{SMOGoriginal}, a measure estimating the grade level required for understanding the text. The SMOG grade is computed as:
\begin{align*}
    \mbox{SMOG} = 1.0430 \sqrt{\frac{30n_{w*}}{n_s}}+3.1291,
\end{align*}
where $n_{w*}$ is the number of words of 3 or more syllables, and $n_s$ is the number of sentences \cite{SMOGoriginal}.

Note that the study showing correlation between noun and verb phrases, and readability, is done on English texts. The SMOG grade as well is defined with the purpose of evaluating English texts. Hence, one must be careful when basing conclusion on these measures when used on Danish. However, we believe they can still provide information about the development, even if the exact computed value might be hard to interpret.

\section{Experiments and Results}\label{sec:results}
In this section, we present the data, the experimental setup, and the results obtained. \secref{data} presents the data, how it is preprocessed and split for training and analysis and some basic statistics. \secref{exp-network} describes the training of the Siamese neural network, while \secref{exp-clustering} describes the  clustering and shows the resulting clusters.

\subsection{Data}\label{sec:data}
The full data set made available to us by MaCom contains around 130K essays by approximately 10K students, with an average length of about 6K characters. The data set was cleaned by removing very short ($\leq 400$) and very long ($\geq 30,000$) texts, in order to get rid of outliers/invalid essays (blank hand-ins, garbled texts, etc.). Furthermore, proper pronouns were substituted with placeholder tokens and the first 200 characters of each text were removed, in an effort to remove any data identifying the real author of the text, as such clues could be picked up by the neural network and lead to overfitting. Finally, authors with less than 5 texts were removed.
Following this cleaning, the data set contains a total of 131,095 Danish essays, written by 10095 authors, with an average 13.0 texts per author, and an average text length of 5894.8 characters.

We partition the clean data into two author disjoint sets: \networkset\ used for training the neural network, and \analset, which we analyze using the trained similarity function. \networkset\ is further split into a training set \trainset\ and a validation set \valset\ (also author disjoint), used for early stopping when training the network. As the analysis relies heavily on a strong similarity function, the majority of the data (around two thirds) is used for \networkset. The exact sizes are given in \tabref{data}.

\begin{table}[h!]
  \centering
  \begin{tabular}{|c|c|c|c|c|}
    \hline
    Data set    & \#students & \#texts & \#\Sim \\
    \hline
    \trainset & 5418        & 70432      & 934720 \\
    \valset   & 989         & 12997      & 173536 \\
    \analset  & 3688        & 47666      & N/A \\
    \hline
    Total     & 10095       & 131,095    & 1108256 \\
    \hline
  \end{tabular}
  \caption{Data set overview. The table lists the number of students and texts, as well as the number of problem instances \#\Sim\ for training the Siamese neural network.
  }\label{tab:data}
\end{table}

\subsubsection*{Data for network training}
For training and evaluating the Siamese neural network, we require problem instances consisting of a pair of texts, and a label indicating whether they are by the same author (positive sample) or by different authors (negative sample). We refer to these instances as \Sim-instances, and generate them for the training set $\trainset$ and the validation set $\valset$.

Positive \Sim-instances are generated by using $\txt_i,\txt_j\in\txtsstud$ with $i\neq j$, while negative instances are generated by using $\txt_i\in\txtsstudi{\otherstudent_1}$ and $\txt_j\in\txtsstudi{\otherstudent_2}$, where $i,j,\otherstudent_1,\otherstudent_2$ are selected at random, with $\otherstudent_1 \neq \otherstudent_2$. A balanced 50:50 data set is generated by generating the maximum number of positive instances for each student, and an equal number of negative instances.
The final numbers of \Sim-instances for \trainset\ and \valset\ are shown in \tabref{data}.

Note, that in generating these samples, we assume all claimed authors in the data to be the real authors; in reality, several students may use ghostwriters or plagiarism, in which case the labels will be wrong. However, we expect that the number of invalid labels is low.

\subsubsection*{Data for clustering and analysis}
\begin{figure}
        \begin{tikzpicture}
\begin{axis}[
width=0.5\columnwidth,
xmin=6,xmax=41,
ymin=0,
xlabel={\small{\#texts}},
ylabel={\small{\#students}},
]
\addplot[mark=none, color=colDefault, fill=colDefaultFill, ybar interval] table [x=texts, y=students, col sep=semicolon]{data/handindist.csv};
\end{axis}
\end{tikzpicture}%
%
        \begin{tikzpicture}
\begin{axis}[
width=0.5\columnwidth,
xmin=0,xmax=36,
ymin=0,
xlabel={\small{Months}},
ylabel={\small{\#texts}},
]
\addplot[mark=none, color=colDefault, fill=colDefaultFill, ybar interval] table [x=month, y=count, col sep=semicolon]{data/histogram.csv};
\end{axis}
\end{tikzpicture}%
%
    \caption{Statistics for \analset. Distribution of students according to number of essays written (left) and total number of essays written at any time during a students stay in high school (right).}
    \label{fig:data-stats}
\end{figure}
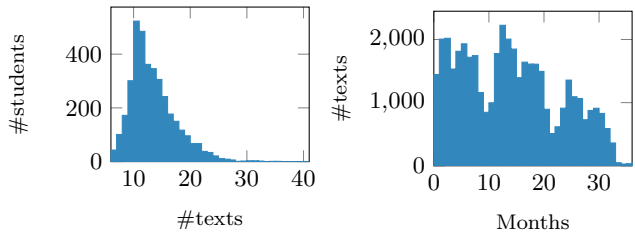

The clustering is performed on the remaining data in \analset. Each data point consists of a single student and their texts. As mentioned, an author has around 13 texts in average, distributed over three years; the actual distribution is shown in \figref{data-stats} (left)\footnote{Recall that, students with less than 5 essays is not considered in this study.}.

\figref{data-stats} (right) shows the number of essays handed in during the three years of high school. The summer vacations are clearly visible in the plot. Note also, that the number of hand-ins drops during the third year. A few students spend more than three years (not shown in the figure), but as only a few students hand-in after 30 months, we consider only the data within 30 months in the experiments\footnote{The time span considered is smaller than three years (36 months), since we measure the time from first hand-in until the last. Combining this with vacation and finals, most students appear to only be active within the 30 month period.}.

\subsection{Neural Network Training}\label{sec:exp-network}
\begin{figure}[t]
    \centering
    \begin{tikzpicture}
  \begin{axis}[
      width=0.47\linewidth,
      height=0.47\linewidth,
      xmin=1,xmax=6,
      xtick=data,
      xlabel={Epoch},
      ylabel={Loss},
      ylabel style={
      },
      yticklabel style={
        /pgf/number format/fixed,
        /pgf/number format/precision=3
      },
      scaled y ticks=false,
    ]
    \addplot[mark=none, color=colDefault, line width=1.5pt] table [x=Epoch, y=TrainLoss]{data/log.txt};
    \addplot[mark=none, color=colDefault, line width=1.5pt, dashed] table [x=Epoch, y=ValLoss]{data/log.txt};
  \end{axis}
\end{tikzpicture}
\begin{tikzpicture}
  \begin{axis}[
      width=0.47\linewidth,
      height=0.47\linewidth,
      xmin=1,xmax=6,
      xtick=data,
      xlabel={Epoch},
      ylabel={Accuracy},
      ylabel style={
      },
      yticklabel style={
        /pgf/number format/fixed,
        /pgf/number format/precision=3
      },
      scaled y ticks=false,
    ]
    \addplot[mark=none, color=colCluster2, line width=1.5pt] table [x=Epoch, y=TrainAcc]{data/log.txt};
    \addplot[mark=none, color=colCluster2, line width=1.5pt, dashed] table [x=Epoch, y=ValAcc]{data/log.txt};
  \end{axis}
\end{tikzpicture}
    \caption{Plot of training (solid) and validation (dashed) loss (left) and accuracy (right), the latter computed by rounding the output. Minimum validation loss was obtained at epoch 5.}
    \label{fig:training}
\end{figure}
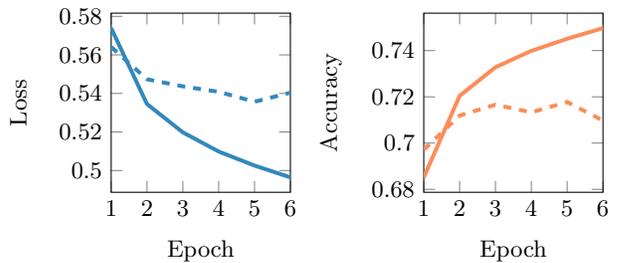

The similarity network described in \secref{network} was implemented using TensorFlow. We generate \Sim-instances for \trainset\ and \valset, and optimize the network for cross entropy using the Adam optimizer. 
The final network obtains a training loss of $0.5026$ and a validation loss of $0.5357$. Rounding the computed similarity to 0 or 1, we can compute an accuracy of $0.7451$ for the training set and an accuracy of $0.7178$ for the validation set. \figref{training} shows a plot of the loss and accuracy, as the network was trained.

\subsection{Clustering}\label{sec:exp-clustering}
Using the similarity network to compute the similarity function \simfuncname, we construct profiles as described in \secref{clustering}. We found that using $m=2$ texts for determining the initial writing style yielded good results. Thus, profile $\profile{\student}$ consists of $|\txtsstud|-1$ pairs $(\ptime_j, \psim_j)$ with:
\begin{align*}
    \psim_j = \frac{\simfunc{\txt_{j+1}}{\txt_1}+\simfunc{\txt_{j+1}}{\txt_2}}{2}.
\end{align*}
and $\ptime_j$ being the time in months since hand-in of $\txt_2$.
With a single profile constructed per student, the total number of profiles is equal to the number of students, as given in \tabref{data}.
As mentioned, the lengths of the profiles depend on the number of texts written during the time, they spend in high school. Thus the distribution of the lengths of profiles follows that presented in \figref{data-stats} (left).

\begin{figure}[t]
    \centering
    \begin{tikzpicture}
  \begin{axis}[
      width=0.6\linewidth,
      height=0.6\linewidth,
      xmin=2,xmax=9,
      xlabel={$k$},
      ylabel={\clustererror},
      ylabel style={
      },
      yticklabel style={
        /pgf/number format/fixed,
        /pgf/number format/precision=3
      },
      scaled y ticks=false,
    ]
    \addplot[mark=none, color=colDefault, line width=1.5pt] table [x=k, y=err, col sep=semicolon]{data/select.csv};
  \end{axis}
\end{tikzpicture}
    \caption{The cluster error \clustererror\ obtained for various values of $k$.}
    \label{fig:select-k}
\end{figure}
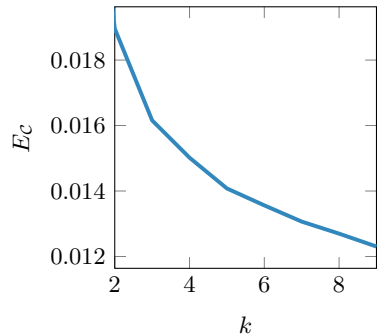

We now apply the elbow method in order to determine the optimal number of clusters $k$. We compute a clustering for $k=2,3,...,9$, and plot the resulting cluster error \clustererror\ in \figref{select-k}. 

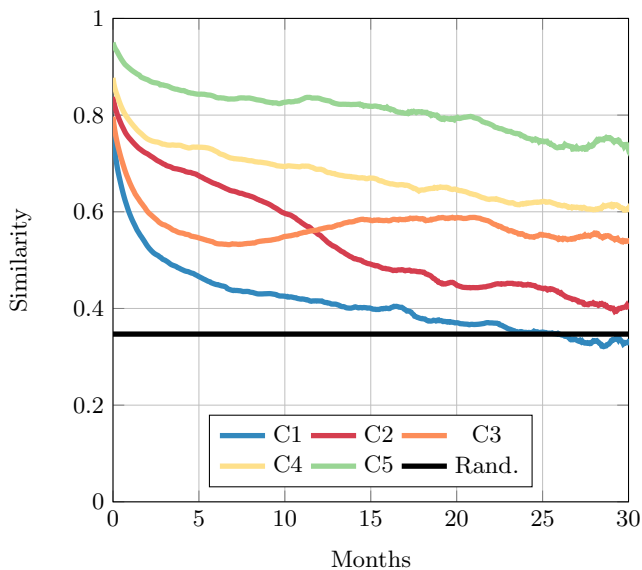
\begin{figure}[t]
    \centering
    \begin{tikzpicture}
\begin{axis}[FullClusterPlot]
	\addplot[mark=none, color=colCluster0, line width=2pt] table [x=idx, y=sim, col sep=semicolon]{data/5-l2-c0.csv};
	\addlegendentry{C1}

	\addplot[mark=none, color=colCluster1, line width=2pt] table [x=idx, y=sim, col sep=semicolon]{data/5-l2-c1.csv};
	\addlegendentry{C2}

	\addplot[mark=none, color=colCluster2, line width=2pt] table [x=idx, y=sim, col sep=semicolon]{data/5-l2-c2.csv};
	\addlegendentry{C3}

	\addplot[mark=none, color=colCluster3, line width=2pt] table [x=idx, y=sim, col sep=semicolon]{data/5-l2-c3.csv};
	\addlegendentry{C4}

	\addplot[mark=none, color=colCluster4, line width=2pt] table [x=idx, y=sim, col sep=semicolon]{data/5-l2-c4.csv};
	\addlegendentry{C5}

	\addplot[mark=none, color=colClusterRandom] coordinates {(0,0.34699403107089527) (30,0.34699403107089527)};
	\addlegendentry{Rand.}
\end{axis}
\end{tikzpicture}
    \caption{The curves representing the five clusters found. The average similarity between random texts by different students is also plotted.}
    \label{fig:all-cluster}
\end{figure}

Based on \figref{select-k}, we select $k=5$, as the curve flattens considerable for $k=6$. The final clustering is performed, obtaining five clusters: \cluster{1}, \cluster{2}, \cluster{3}, \cluster{4}, and \cluster{5}, with a cluster error of $\clustererror = 0.01407$. The curves representing the final clusters are shown in \figref{all-cluster}, while \tabref{cluster-member-count} lists the number of members in each cluster. Furthermore, we sampled two million random pairs of texts with random (different) authors, and computed the similarity for these samples, obtaining an average of $0.3470$. This average is also plotted in \figref{all-cluster}, while the similarity with respect to time is plotted as a heat map in \figref{rand-sim}.
\begin{table}[h]
    \centering
    \begin{tabular}{|c|c|}
        \hline
        Cluster & \#students \\
        \hline
        \cluster{1} & 603 \\
        \cluster{2} & 720 \\
        \cluster{3} & 884 \\
        \cluster{4} & 969 \\
        \cluster{5} & 512 \\ 
        \hline
    \end{tabular}
    \caption{The number of students in each cluster.}
    \label{tab:cluster-member-count}
\end{table}

\begin{sidewaysfigure*}
    \resizebox{\textwidth}{!}{
        \begin{tikzpicture}
\begin{groupplot}[
group style={
group size=5 by 4,
x descriptions at=edge bottom,
y descriptions at=edge left,
vertical sep=10pt,
horizontal sep=10pt,
},
]
\nextgroupplot[ClusterPlot, title=\cluster{1},ylabel=Similarity]
\addplot[Cluster, color=colCluster] table [x=idx, y=sim, col sep=semicolon]{data/5-l2-c0.csv};
\addplot[ClusterError, color=colCluster,name path=upper] table [x=idx, y=upper, col sep=semicolon]{data/5-l2-c0.csv};
\addplot[ClusterError, color=colCluster,name path=lower] table [x=idx, y=lower, col sep=semicolon]{data/5-l2-c0.csv};
\addplot[ClusterError,color=colCluster] fill between[of = upper and lower];
\nextgroupplot[ClusterPlot, title=\cluster{2}]
\addplot[Cluster, color=colCluster] table [x=idx, y=sim, col sep=semicolon]{data/5-l2-c1.csv};
\addplot[ClusterError, color=colCluster,name path=upper] table [x=idx, y=upper, col sep=semicolon]{data/5-l2-c1.csv};
\addplot[ClusterError, color=colCluster,name path=lower] table [x=idx, y=lower, col sep=semicolon]{data/5-l2-c1.csv};
\addplot[ClusterError,color=colCluster] fill between[of = upper and lower];
\nextgroupplot[ClusterPlot, title=\cluster{3}]
\addplot[Cluster, color=colCluster] table [x=idx, y=sim, col sep=semicolon]{data/5-l2-c2.csv};
\addplot[ClusterError, color=colCluster,name path=upper] table [x=idx, y=upper, col sep=semicolon]{data/5-l2-c2.csv};
\addplot[ClusterError, color=colCluster,name path=lower] table [x=idx, y=lower, col sep=semicolon]{data/5-l2-c2.csv};
\addplot[ClusterError,color=colCluster] fill between[of = upper and lower];
\nextgroupplot[ClusterPlot, title=\cluster{4}]
\addplot[Cluster, color=colCluster] table [x=idx, y=sim, col sep=semicolon]{data/5-l2-c3.csv};
\addplot[ClusterError, color=colCluster,name path=upper] table [x=idx, y=upper, col sep=semicolon]{data/5-l2-c3.csv};
\addplot[ClusterError, color=colCluster,name path=lower] table [x=idx, y=lower, col sep=semicolon]{data/5-l2-c3.csv};
\addplot[ClusterError,color=colCluster] fill between[of = upper and lower];
\nextgroupplot[ClusterPlot, title=\cluster{5}]
\addplot[Cluster, color=colCluster] table [x=idx, y=sim, col sep=semicolon]{data/5-l2-c4.csv};
\addplot[ClusterError, color=colCluster,name path=upper] table [x=idx, y=upper, col sep=semicolon]{data/5-l2-c4.csv};
\addplot[ClusterError, color=colCluster,name path=lower] table [x=idx, y=lower, col sep=semicolon]{data/5-l2-c4.csv};
\addplot[ClusterError,color=colCluster] fill between[of = upper and lower];
\nextgroupplot[ClusterPlotVocab,ylabel=SMOG]
\addplot[Data, color=colCluster] table [x=idx, y=smogs, col sep=semicolon]{data/5-l2-c0.csv};
\nextgroupplot[ClusterPlotVocab]
\addplot[Data, color=colCluster] table [x=idx, y=smogs, col sep=semicolon]{data/5-l2-c1.csv};
\nextgroupplot[ClusterPlotVocab]
\addplot[Data, color=colCluster] table [x=idx, y=smogs, col sep=semicolon]{data/5-l2-c2.csv};
\nextgroupplot[ClusterPlotVocab]
\addplot[Data, color=colCluster] table [x=idx, y=smogs, col sep=semicolon]{data/5-l2-c3.csv};
\nextgroupplot[ClusterPlotVocab]
\addplot[Data, color=colCluster] table [x=idx, y=smogs, col sep=semicolon]{data/5-l2-c4.csv};
\nextgroupplot[ClusterPlotPhrases,ylabel=Phrases]
\addplot[Noun, color=colCluster] table [x=idx, y=nouns, col sep=semicolon]{data/5-l2-c0.csv};
\addplot[Verb, color=colCluster] table [x=idx, y=verbs, col sep=semicolon]{data/5-l2-c0.csv};
\nextgroupplot[ClusterPlotPhrases]
\addplot[Noun, color=colCluster] table [x=idx, y=nouns, col sep=semicolon]{data/5-l2-c1.csv};
\addplot[Verb, color=colCluster] table [x=idx, y=verbs, col sep=semicolon]{data/5-l2-c1.csv};
\nextgroupplot[ClusterPlotPhrases]
\addplot[Noun, color=colCluster] table [x=idx, y=nouns, col sep=semicolon]{data/5-l2-c2.csv};
\addplot[Verb, color=colCluster] table [x=idx, y=verbs, col sep=semicolon]{data/5-l2-c2.csv};
\nextgroupplot[ClusterPlotPhrases]
\addplot[Noun, color=colCluster] table [x=idx, y=nouns, col sep=semicolon]{data/5-l2-c3.csv};
\addplot[Verb, color=colCluster] table [x=idx, y=verbs, col sep=semicolon]{data/5-l2-c3.csv};
\nextgroupplot[ClusterPlotPhrases]
\addplot[Noun, color=colCluster] table [x=idx, y=nouns, col sep=semicolon]{data/5-l2-c4.csv};
\addplot[Verb, color=colCluster] table [x=idx, y=verbs, col sep=semicolon]{data/5-l2-c4.csv};
\nextgroupplot[ClusterPlotWords,ylabel=\#Word]
\addplot[Data, color=colCluster] table [x=idx, y=words, col sep=semicolon]{data/5-l2-c0.csv};
\nextgroupplot[ClusterPlotWords]
\addplot[Data, color=colCluster] table [x=idx, y=words, col sep=semicolon]{data/5-l2-c1.csv};
\nextgroupplot[ClusterPlotWords]
\addplot[Data, color=colCluster] table [x=idx, y=words, col sep=semicolon]{data/5-l2-c2.csv};
\nextgroupplot[ClusterPlotWords]
\addplot[Data, color=colCluster] table [x=idx, y=words, col sep=semicolon]{data/5-l2-c3.csv};
\nextgroupplot[ClusterPlotWords]
\addplot[Data, color=colCluster] table [x=idx, y=words, col sep=semicolon]{data/5-l2-c4.csv};
\end{groupplot}
\end{tikzpicture}
    }
    \caption{Detailed plots of the five clusters found. The top plot shows the similarity curve, with the middle 90\% of profiles in each cluster plotted as well. The second plot from the top shows the development of the SMOG grade, while the third plot shows noun (solid) and verb (dashed) phrases. Finally the bottom plot shows the development of number of words written.}
    \label{fig:cluster-detail}
\end{sidewaysfigure*}
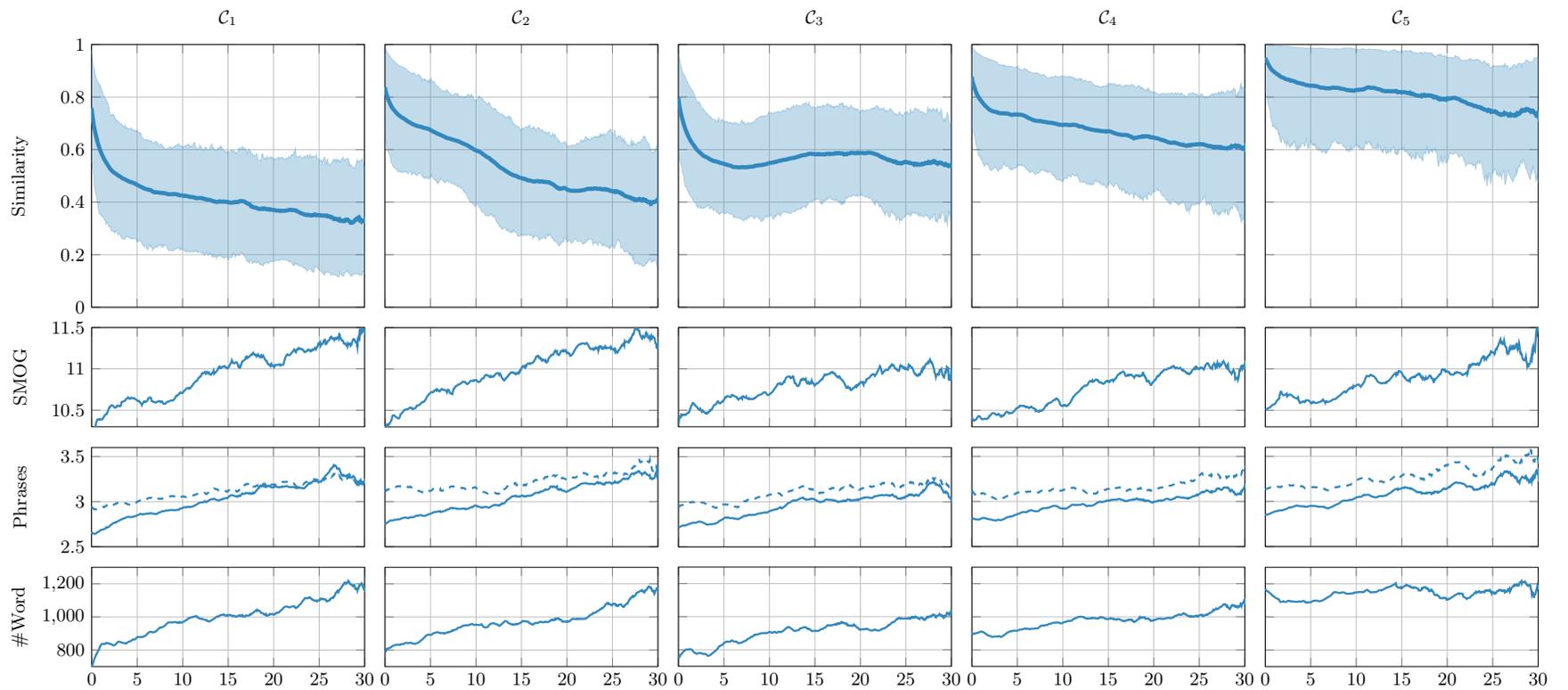

Finally, \figref{cluster-detail} shows a more detailed view for each cluster. The similarity curve plots include a plot of the middle 90\% of profiles in each cluster. The SMOG score, the noun and verb phrases, and the average text length (in words) are also plotted, as indicators for writing quality changes for the given cluster, see also \secref{clustering}.

Note, that in visualizing and inspecting the clusters, we consider only data until 30 months, since, as mentioned, only few students are active after 30 months, and the number of data points contributing to that part of the cluster curve becomes small.

\section{Analysis and Discussion}\label{sec:analysis}
This section presents our analysis and discussion of the five clusters found. \secref{anal-cluster} describes  and discusses the characteristics of each cluster. \secref{anal-random} discussed how similarity between random students behaves with time.

\subsection{Cluster Analysis}\label{sec:anal-cluster}
When analyzing the clusters, three properties are important in order to understand a cluster:
the initial value of the curve, the shape of the curve, and the total change from start to end. 
The \emph{initial value of the curve}
describes the similarity between the second text of a student and their initial writing style, which is based on the first two texts of the student. Thus a smaller initial value indicates
a high initial variance in writing style, which could be an indication of a developing writing style.
The \emph{shape of the curve} describes the rate of change
in writing style.
And finally the \emph{total change} tells us how much the writing style has evolved.

While the similarity curves themselves give no information about \emph{the quality} of the writing, we will use the indicators of writing style, described in \secref{method}, in the discussion: the SMOG grade, and the noun and verb phrases per sentences. For each of these indicators, the average curves for each cluster are plotted in \figref{cluster-detail}.

Before discussing each cluster in detail, we note some patterns common for all clusters. Across all clusters, it seems the number of words written increases (with the exception of \cluster{5}), and the increase seems to be correlated with the corresponding decrease in similarity. Furthermore, on average, students in all clusters appear to be improving with respect to the quality metrics. While positive, some clusters see a smaller increase than others, indicating that these clusters represents suboptimal development profiles.
Finally, we note for the SMOG grade, that the maximum increase, occurring in \cluster{1}, is only slightly above 1, which might not seem impressive across three years. However, as discussed in \secref{clustering}, the SMOG grade is a measure designed for readability of English texts, and thus may not be entirely accurate for Danish texts.

Below follow detailed descriptions of each cluster:
\begin{enumerate}
    \item[\cluster{1}]
    The initial similarity of \cluster{1} is the lowest among the clusters found. Furthermore, the similarity drops rapidly during the first year, and continues the decline, leading to \cluster{1} having the lowest final similarity with the initial writing style, among all the clusters. In fact, 
    the similarity between the first and the last assignments
    for students in this cluster
    is so low, that they could just as well have been written by different students,
    as can be seen when comparing to the average similarity between random students plotted in \figref{all-cluster}.
    Thus, \cluster{1} contains students with a significant change in writing style, happening mostly during the first year of high school.
    Considering the other metrics plotted in \figref{cluster-detail}, we first note the increase in
    number of words written, as it is particular extreme in the case of \cluster{1}, 
    increasing by almost a factor 2 from start till end.
    This increase also helps
    explain the decrease in similarity in two ways: a) length is itself a part of writing style recognized by the network, and b) it seems that writing style changes is correlated with when you start writing more.
    Looking at the SMOG grade, we see an overall large increase, indicating that the students of \cluster{1} does indeed improve, especially compared to the other clusters.
    Nouns and verbs per sentence are also both increasing, which also indicates that the students in this cluster write longer sentences.
    \item[\cluster{2}]
    The initial value of cluster \cluster{2}\ of about 0.84 (third highest among the clusters) indicates an initial low variance in writing style, but the following drop
    to about 0.4 is quite significant,
    indicating that the writing style of students in \cluster{2}\ change a lot during high school, similar to \cluster{1}.
    However, where \cluster{1} had a sudden drop in similarity, the change in \cluster{2} is more constant.
    
    Considering the other metrics, we see the number of words written is increasing from about 800 to almost 1200, while the SMOG grade is again showing a large increase from about 10.3 to 11.3. Noun and verb phrases see modest increases.
    All in all, the metrics indicate a good development of writing style among students in \cluster{2}, similar to \cluster{1}. However, the more gradual change in similarity of \cluster{2} is preferable to that of \cluster{1}, as the development does not stagnate already after the first year.
    %
    %
    \item[\cluster{3}]
    After a significant initial drop in similarity in \cluster{3},
    the similarity actually increases again
    after the first year, showing the students in this cluster actually reverts to
    writing style more similar to their original work, before dropping a bit again in the last months.
    This corresponds well with a smaller improvement in e.g. SMOG grade (around 0.5) compared to the other clusters.
    
    The setback seems to start around the first summer vacation.
    While not necessarily bad (as students could be reverting back from a worsened writing style), the increase in similarity could indicate reverting to a worse writing style. As such, students in \cluster{3} may be at risk.
    Many remedies for helping these students could be imagined, from simply encouraging the student to write during their vacation, to going to summer school.
    \item[\cluster{4}]
    The similarity of \cluster{4} drops slightly at first, but then decreases slowly at a constant pace,
    until it reaches a similarity of about 0.6. The total change is smaller than several of the other clusters, as is the improvement in both SMOG grade and noun/verb phrases, indicating that students in the cluster improve less than students in e.g. \cluster{1} and \cluster{2}.
    
    This indicates suboptimal development among students in \cluster{4}; while we do not see students reverting back, as in \cluster{3}, the lower increase in SMOG grade is alarming, indicating students in this cluster may be at risk, and in need of extra attention or encouragement.
    As for \cluster{3}, the total number of words also increases only slightly at a steady pace, from around 900 to 1100.
    %
    \item[\cluster{5}]
    Cluster \cluster{5} seems quite distinct from the other clusters. Most notably, students in this cluster
    have the highest initial similarity, while also decreasing the least amount, ending with a very high similarity of about 0.75.
    Furthermore, the number of words written is quite high and remains fairly stable, which is quite different from the other clusters, and might be part of the reason the decrease in similarity is as low as it is.
    Despite the fall in similarity being so low, we still see an increase in SMOG score from about 10.5 to a bit below 11.5,
    indicating that students are, in fact, improving. A similar pattern occurs for the noun and verb phrases. 
    
    The higher-than-average initial SMOG grade and number of words written, indicates that students in \cluster{5} are the initially strong students. While they do develop their writing style, they do not improve as much as students in \cluster{1} and \cluster{2}; this could be an indication, that schools do not manage to properly encourage/teach students, who are initially strong. 
\end{enumerate}
While not included in the plots, we also investigated several other metrics for the clusters and the set of students in general. Most notably, the average word length increases with time for all clusters. A similar trend was seen in \cite{changeovertime}, although the study was in a very different setting and time frame.

Summarizing the clusters, the development in SMOG grade was greatest for \cluster{1} and \cluster{2}, making those clusters appear the most beneficial for writing style development. While students in \cluster{5} also increased their SMOG grade, they started higher than the students in the other clusters, and did not manage to improve as much as \cluster{1} and \cluster{2}.
As to \cluster{3} and \cluster{4}, they seem to be suboptimal with regards to writing style development,
and students in these clusters may need attention.

Looking at \tabref{cluster-member-count}, we see that \cluster{3} and \cluster{4} are the largest individually, indicating that quite a few students are exhibiting suboptimal writing style development. However, the majority of students included in our data are located in \cluster{1}, \cluster{2} and \cluster{5}, indicating optimal or at least fair development through high school.

\subsection{Investigating Similarity Between Random Students}\label{sec:anal-random}
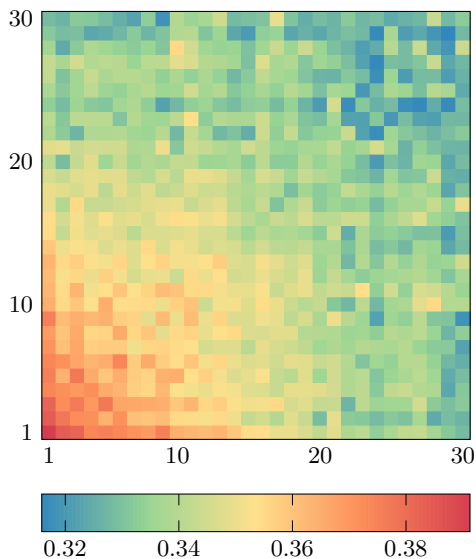
\begin{figure}
    \centering
    \begin{tikzpicture}
  \begin{axis}[
      width=\linewidth,
      unit vector ratio=1 1 1,
      colorbar,
      colorbar horizontal,
      xtick={0,9,19,29},
      xticklabels={1,10,20,30},
      ytick={0,9,19,29},
      yticklabels={1,10,20,30},
      major tick length=0,
      enlargelimits=false,
      axis on top,
      xmin=-0.5,xmax=29.5,
      ymin=-0.5,ymax=29.5,
      colormap name=simcolormap,
      colorbar style={
        xtick style={
          color=black
        }
      },
    ]
    \addplot [matrix plot*,point meta=explicit] file [meta=index 2] {data/simmap.dat};
  \end{axis}
\end{tikzpicture}
    \caption{Heat map showing the average similarity between different authors, depending on how long time the two authors have been in high school.}
    \label{fig:rand-sim}
\end{figure}

As mentioned, we also investigated how the similarity develops between different students, across the time spent in high school. Based on roughly 2 million sampled text pairs from different students, we computed the average similarity between random students to be $0.3470$. As seen in \figref{all-cluster}, the similarity observed among students in \cluster{1} actually drops below this value. This motivated a further investigation of how the similarity between different authors behave on average, conditioned on how long time they each have spend in high school. Based on the samples, we constructed the heat map shown in \figref{rand-sim}.
 
 The plot shows students starting out similar in writing style and then becoming less similar as time passes. The most surprising thing to notice is that a student in their first year and a student in their third year are equally or even more similar in writing style on average, compared to two different students in their third year. One explanation could be that the initial space of possible writing styles start out small and grows as students are educated,
 i.e. writing styles among students coming from primary school are fairly similar, but grow more diverse during high school. 
 One would expect some writing styles to diminish or even disappear, but from this data it looks like more new and diverse writing styles develop, than disappear. And not only that; the amount of possible directions for the writing style to develop is so large, that we see first and third year students as equally or more similar on average, than two students both within their third year. 
 
 Education is sometimes accused of destroying individuality and/or creativity; these findings indicate the opposite to such claims, at least in regards to writing style.
 
\newpage
\section{Conclusions and Future Work}
We trained a Siamese neural network to be able to tell people apart by their writing, and used this network as a similarity function for analyzing the development of writing style in Danish high schools. Writing style development profiles were constructed for 3688 students, and five clusters were found and discussed. Based on quality indicated by noun/verb phrases and SMOG grade, two were found to be optimal, while three were found to be suboptimal, especially two clusters exhibited limited improvement.

The optimal clusters both exhibited a large degree of change in writing style, although at different rates, while the suboptimal clusters showed less development, with one cluster even reverting back to an earlier writing style.
The setback in similarity occurred around the summer vacation after the first year. The effect of summer vacation on student learning is highly discussed topic among researchers, teachers, and parents \cite{summer1}; in the case of the found cluster, the effect appears to be negative.

One tendency, we saw in all clusters, was that writing style changed more when students start writing more words in their essays. It does not seem surprising that your writing style changes as you write more, but it could be an  indication of even more: writing style changes, when students are pushed out of their comfort zone, i.e. in the end of their assignments, when they write more than what they usually do. It could be interesting to investigate the scenario, where a student starts writing longer texts: does changes in writing style occur in the entire text, or only near the end, where the student is literally writing more than before?

Furthermore, we saw from \figref{rand-sim} how students become less alike, as they go through high school. Specifically, we saw how first year and third year students had higher or equal writing style similarity than two students both in third year, indicating that as Danish students go through high school, their writing styles diverge and become more individual.


\subsection{Future Work}
It is easy to pose several new questions based on the clusters found and the conclusions made above.

With regards to improving the analysis, using different quality measures tailored to Danish instead of SMOG would be interesting. Another way would be to consider the grades given to the students (as many essays in Danish high school are graded individually), although good writing style is only a requirement for a good grade, but not sufficient.

As mentioned above, one could also consider a more fine grained analysis, by investigating style changes within texts, and maybe even being able to pinpoint exactly where in a text the writing style develops/changes. One could easily imagine drawing inspiration from studies of style breach detection \cite{stylebreach1,stylebreach2,stylebreach3}.

One could also investigate prediction of writing style development, possibly based on the methods used in this study. This would allow for an early warning system, allowing identification of at-risk students, e.g. students likely to have a setback in writing style due to summer vacation.

The methods used in this study build upon methods used for authorship verification, in which Siamese networks are utilized directly in order to verify authorship \cite{ESANN19Ghostwriter}. While a sudden deviation in writing style could be an indication of a ghost writer, detecting these reliably using our method will probably not be able to compete with the more direct methods. However, the results obtained here could potentially be used to improve authorship verification techniques, with respect to the fairness perspective:
The fact, that the clusters found show such different similarity development, is of interest from a fairness perspective. Fairness is a general issue in machine learning algorithms where the predictions have severe consequences \cite{fairness1,fairness2}. In the setting of ghost writing detection in high school it is extremely difficult to get non-artificial negative samples and even guaranteeing correctness of labels is rare in large scale data sets. Which makes fairness even more difficult to measure than usual. It could be interesting to check that clusters such as \cluster{5}, which would be the cluster most likely to be classified as a false negative, have a representative distribution in regards to gender, race, social status, etc.

Another interesting course of study, would be to further investigate the fact that students seem to become less similar during high school. It could be interesting to pursue this on a larger timescale, perhaps all the way from primary school and on through college. 
Another take could be to look at how similarity in writing among people behaves with age after they have finished their education. Will the trend continue?

Finally, one could investigate how similarity in writing develops among the genders. Several studies have shown, with some success, that gender can be predicted from writing \cite{gender,ageandgender2,ageandgender}, but no one has settled whether this is due to biology or environment. One could try to answer this question by looking at how similarity in writing style  changes 
with age, while considering three groups: female-female, male-male, female-male.
If the cross gender similarity changes faster than same gender similarity, it would be an indication that the differences in writing style are taught, more than it is something you are born with.

\section{Acknowledgments}
The work is supported by the Innovation Fund Denmark through the Danish Center for Big Data Analytics Driven Innovation (DABAI) project. The authors would like to thank MaCom for the cooperation.


\bibliographystyle{plain}
\bibliography{bibl}  
\end{document}